\documentclass{rmf-d}
\usepackage{nopageno, rmfbib, multicol,times,epsf,amsmath,amssymb,cite}
\usepackage[latin1]{inputenc}
\usepackage[]{caption2}
\usepackage{graphics}
\usepackage[demo]{graphicx}
\usepackage{hyperref}
\usepackage{lipsum}
\usepackage{wrapfig, blindtext}
\usepackage{comment}
\usepackage{tabularx}
\usepackage{float}
\usepackage{subfigure}
\usepackage{braket}
\usepackage{mathtools}

\def\rmfcornisa{
}

\newcommand{\MSb}{{\overline{\rm MS}}}
\def\rmfcintilla{
}
\clearpage \rmfcaptionstyle 
\begin{document}
\title{Proton generalized parton distributions from lattice QCD
\vspace{-6pt}}
\author{Aurora Scapellato$^1$, Constantia Alexandrou$^{2,3}$, Krzysztof Cichy$^4$,\\ Martha Constantinou$^1$, Kyriakos Hadjiyiannakou$^3$, Karl Jansen$^5$, Fernanda Steffens$^6$}
\address{
\vskip 0.25cm
$^1$Department of Physics,  Temple University,  Philadelphia,  PA 19122 - 1801, USA   \\
 \vskip 0.05cm
$^2$Department of Physics, University of Cyprus,  P.O. Box 20537,  1678 Nicosia, Cyprus\\
  \vskip 0.05cm
$^3$Computation-based Science and Technology Research Center,
  The Cyprus Institute, 20 Kavafi Str., Nicosia 2121, Cyprus \\
  \vskip 0.05cm
$^4$Faculty of Physics, Adam Mickiewicz University, Uniwersytetu Pozna\'nskiego 2, 61-614 Pozna\'{n}, Poland  \\
  \vskip 0.05cm
$^5$NIC, DESY,
  Platanenallee 6,
  D-15738 Zeuthen,
  Germany   \\
  \vskip 0.05cm
$^6$Institut f\"ur Strahlen- und Kernphysik, Rheinische
  Friedrich-Wilhelms-Universit\"at Bonn, Nussallee 14-16, 53115 Bonn
}

\maketitle
\begin{abstract}
\vspace{1em} Momentum and spatial distributions of quarks and gluons inside hadrons are typically encoded in the so-called generalized parton distributions (GPDs). GPDs are  multi-dimensional quantities that are very challenging to extract, both experimentally and within lattice QCD. We present the first lattice results on the $x$-dependence of isovector unpolarized, helicity and transversity GPDs of the proton, obtained from lattice QCD using an ensemble of $N_f=2+1+1$ maximally twisted mass fermions, with pion mass $M_\pi=260$ MeV and lattice spacing $a\simeq 0.093$ fm. Our calculations use the quasi-distribution formalism and the final distributions are presented in the $\MSb$ scheme at a renormalization scale of 2~GeV.   \vspace{1em}
\end{abstract}
\keys{Lattice QCD, nucleon structure, generalized parton distributions  \vspace{-4pt}}
\pacs{   \bf{\textit{11.15.Ha, 12.38.Gc, 12.60.-i, 12.38.Aw}}    \vspace{-4pt}}
\begin{multicols}{2}

\section{Introduction}
The inner structure of hadrons is governed by the highly non-trivial dynamics of the
strong interactions among quarks and gluons, which are collectively called partons. Interactions among partons
are studied, among others, through generalized parton distributions
(GPDs)~\cite{Muller:1994ses,Ji:1996ek,Radyushkin:1996nd,Ji:1996nm}, that are more general functions with respect to form factors (FFs) and parton distribution functions (PDFs). In fact, GPDs parametrize nonforward matrix elements $\langle P'|\dots |P\rangle$ of non-local operators and as such, they depend on three variables: $x$ - the longitudinal momentum fraction of a given parton with respect to the hadron's momentum, $t$ - the square of the four-momentum transferred to the target, i.e. $(P'-P)^2$, and the skewness $\xi$ - the change in the longitudinal
momentum fraction induced by the momentum transfer. While at $\xi\neq 0$ no simple probabilistic interpretation exists, at $\xi= 0$ GPDs describe the probability to find a parton with a longitudinal momentum fraction $x$ and at a given distance from the center of the momentum of the hadron~\cite{Diehl:2005jf}. In addition, at $t=0$ and $\xi=0$ GPDs reduce to PDFs and the $n=0$ Mellin moments, at a given $t$, are the elastic FFs. Thus, GPDs provide a unified picture of the inner structure of hadrons.

A standard way to access GPDs (and many other partonic functions) relies on analyses of high-energy scattering experiments. This is possible due to the asymptotic freedom of the strong interactions at large energies, that allows to formulate factorization theorems for a large class of scattering processes, separating short distance interactions from non-perturbative dynamics (encoded in e.g. the GPDs). It is expected that, in the near future, GPDs will be extracted with unprecedented accuracy by experimental efforts that include the planned Electron-Ion-Collider (EIC)~\cite{AbdulKhalek:2021gbh} and the 12 GeV upgrade program at Jefferson Lab~\cite{Burkert:2018nvj}.

A complementary approach  to experiments relies on the lattice formulation of Quantum Chromodynamics (QCD). However, accessing GPDs within lattice QCD is not straightforward because the relevant matrix elements receive contributions only along the light-cone and light-cone separations do not exist on a Euclidean lattice. The seminal work by X.~Ji on the quasi-distributions~\cite{Ji:2013dva} has opened new perspectives in this field, showing that purely spatial correlations at finite hadron boost can be factorized into the desired distributions through a perturbative matching coefficient, within Large Momentum Effective Theory (LaMET)~\cite{Ji:2014gla,Ji:2020ect}. Since its introduction, in 2013, the quasi-distribution formalism has been extensively studied both theoretically and on the lattice. 
Meanwhile, a number of other approaches have also been proposed, such as the good ``lattice cross sections"~\cite{Ma:2014jla,Ma:2017pxb,Sufian:2020vzb}, the pseudo-PDFs~\cite{Radyushkin:2017cyf,Orginos:2017kos,Joo:2020spy} and the ``OPE without OPE"~\cite{Chambers:2017dov}. In the last decade all these methods have been applied  in particular to PDFs, leading to very encouraging lattice results that can be qualitatively compared with global QCD analyses. For a summary of these approaches and of the lattice results obtained by different collaborations we refer to the reviews~\cite{Cichy:2018mum,Constantinou:2020pek,Ji:2020ect,Cichy:2021lih}. For GPDs, however, lattice studies are at an exploratory stage because the multi-dimensionality of these objects brings a higher level of complexity.

In this manuscript, we report on our first calculation of the proton (twist-2) GPDs using the twisted-mass formulation of QCD. To simplify the lattice calculation, we focus on the $u-d$ isovector flavor structure for which the disconnected diagrams cancel. Results are presented for selected GPDs and for additional results and details we refer to  Refs.~\cite{Alexandrou:2020zbe,Alexandrou:2021bbo,Alexandrou:2021lyf}.

\section{Quasi-GPDs from lattice QCD}
Quasi-GPDs are extracted from matrix elements in which the quark-bilinear operator has a space-like separation, namely
\begin{equation}
h_\mathcal{O}(P_f,P_i,z,\mu)=\langle N(P_f)|\Bar{\psi}(0)\mathcal{O}W(0,z)\psi(z)|N(P_i)\rangle_{\mu_0} ,    
\label{eq:ME}
\end{equation}
where the Wilson line $W$ is in the same direction as the average momentum boost, $\vec{P}=\frac{\vec{P}_i+\vec{P}_f}{2}$. $\mathcal{O}$ gives access to a specific GPD and we use $\gamma_0$ and $\gamma_5\gamma_3$ for unpolarized and helicity, and $\sigma_{31},\sigma_{32}$ for transversity GPDs, having chosen $\vec{P}=(0,0,P_3)$. Throughout the calculation we employ the Breit frame, in which GPDs are defined, and thus $\vec{P}_f=\vec{P}+\frac{\vec{\Delta}}{2}$ and $\vec{P}_i=\vec{P}-\frac{\vec{\Delta}}{2}$, being $\vec{\Delta}=\vec{P}_f-\vec{P}_i$. The nucleon boost and the momentum transfer are related to each other by the skewness parameter, defined as $\xi=-\frac{\Delta_3}{2P_3}$. The matrix elements of Eq.(\ref{eq:ME}), here renormalized non-perturbatively in the RI-MOM scheme at a scale $\mu_0$, are related to the GPDs through continuum decompositions. In Euclidean space and for the vector operator $\gamma_\mu$ we have
\begin{equation}
\begin{split}
\langle N(P_f)|O_{\gamma_\mu}|N(P_i)\rangle=& \langle\langle \gamma_\mu\rangle\rangle F_H(z,P_3,t,\xi)\\& 
-i\frac{\langle\langle\sigma_{\mu\nu}\rangle\rangle}{2m_N}\Delta_\mu F_E(z,P_3,t,\xi),
\label{eq:decomposition}
\end{split}
\end{equation}
where $\langle\langle \Gamma\rangle\rangle\equiv \bar{u}_N(P_f,S')\Gamma u_N(P_i,S)$, $m_N$ is the nucleon mass and $F_H,F_E$ are the matrix elements of $H,E$ GPDs in coordinate space. The corresponding decompositions for the helicity and transversity GPDs can be found in Refs.\cite{Alexandrou:2020zbe,Alexandrou:2021bbo}. 

The $x$-dependence of the GPDs is recovered in two steps. First, we extract the quasi-GPDs by a Fourier-transform of $F_G$ (where $G=H,E,...$) to momentum space
\begin{equation}
    \widetilde{G}(x,\xi,t,\mu_0,P_3)=\int_{-\infty}^{+\infty}dz\, e^{-iP_3xz}\,F_G(z,\xi,t,P_3,\mu_0). 
\end{equation}
Since $F_G$ are obtained at intervals of the lattice spacing, $z/a\in\mathbb{Z}$, the reconstruction of the continuum distribution in $x$ is an ill-posed inverse problem, that does not have a unique solution. To avoid a model-dependent assumption on the light-cone GPDs, we adopt the Backus-Gilbert method proposed for PDFs in Ref.~\cite{Karpie:2019eiq}. In this work, we follow the implementation described in Refs.~\cite{Alexandrou:2020zbe,Alexandrou:2021bbo,Alexandrou:2021lyf}. Finally, given that quasi- and light-cone GPDs only differ in the ultraviolet region~\cite{Ji:2015qla,Xiong:2015nua}, one can match quasi-GPDs to light-cone GPDs via a coefficient determined in perturbation theory. For the matching coefficient, $C_G$, we use the one-loop expressions derived in Ref.~\cite{Liu:2019urm}, which bring the quasi-GPDs in the RI-MOM scheme to the physical-GPDs in the $\MSb$ scheme. $C_G$ depends on the particular GPD, on the renormalization scale $\mu$ in the $\MSb$ scheme and has a more complicated structure for nonzero skewness. At $\xi=0$ it reduces to the kernel of PDFs. 
The leading perturbative corrections (nucleon mass and higher-twist corrections) between quasi- and light-cone GPDs contribute as $O\left (\frac{m_N^2}{P_3^2},\frac{t}{P_3^2},\frac{\Lambda_{QCD}^2}{(xP_3)^2} \right)$ and therefore a very large value of the nucleon boost is desirable. In practice, the value of the boost is limited by an exponential increase of the noise-to-signal ratio and it determines one of the most important systematic uncertainties in the current lattice QCD calculations.
\section{Lattice setup and matrix elements}
\begin{figure}[H]
   \centering
    \includegraphics[width=0.7\linewidth]{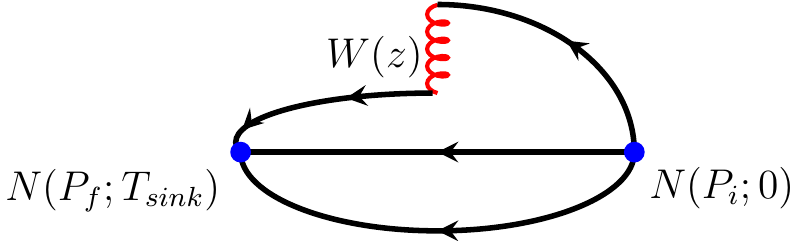}
    \caption{Schematic representation of the (connected) three-point diagram. The initial (source) and final (sink) nucleon states are indicated by $N(P_i;0)$ and $N(P_f;T_{sink})$.}
    \label{fig:diagram}
\end{figure}

We use a gauge ensemble of maximally twisted mass fermions~\cite{ExtendedTwistedMass:2021gbo} with two degenerate light, a strange and a charm quark $(N_f=2+1+1)$, with pion mass $M_\pi~=~260$~MeV, lattice spacing $a\simeq~0.093$~fm and volume $V~=~32^3\times 64$. The matrix elements of Eq.(\ref{eq:ME}) are extracted through appropriate ratios of two- and three-point functions that cancel overlap factors between interpolating fields and nucleon states~\cite{Alexandrou:2020zbe,Alexandrou:2021bbo,Alexandrou:2021lyf}. The correlators are produced by using the momentum smearing technique~\cite{Bali:2016lva} and optimizing the signal for given $\lbrace P_i,P_f\rbrace$. Together with the fact that the Breit frame must be used, for each kinematic setup $\lbrace P_i,P_f\rbrace$ we use a different set of quark propagators. The all-to-all propagators, from the sink positions to insertion points, are computed via the sequential method with the fixed sink approach. The most computationally demanding component of the calculation is depicted by the diagram in Fig.~\ref{fig:diagram}.
\vspace{-0.3cm}
\noindent The source-sink separation is set to $T_{sink}=12a\simeq 1.13$~fm, for which excited states are within $10\%$ of the statistical uncertainties at similar nucleon boosts than the ones used in this work~\cite{Alexandrou:2019lfo}. 
\vspace{0.25cm}

For this first lattice calculation of GPDs we restrict ourselves to one value of the momentum transfer for both zero and nonzero skewness. In particular we have data for $\lbrace t~=-0.69$~GeV$^2,\xi=0\rbrace$ and $\lbrace t=-1.02$~GeV$^2,|\xi|=1/3\rbrace$. At $\xi=0$, we study the momentum dependence of the GPDs using $P_3=0.83,1.25,1.67$~GeV and at $P_3=1.25$~GeV we test the effect of a nonzero $\xi$. In Tab.~\ref{tab:stat_GPDs}, we report the corresponding classes of momenta and the statistics for all operators.
\begin{table}[H]
\hspace*{-3.8cm}
\resizebox{0.91\textwidth}{!}{
\begin{minipage}{\textwidth}
\begin{center}
\renewcommand{\arraystretch}{1.2}
\begin{tabular}{|cccc|cc|}
\hline
$P_3$ [GeV] & $\quad \vec{\Delta}$ $[\frac{2\pi}{L}]\quad$ & $-t$ [GeV$^2$] & $\xi$ & $N_{\rm confs}$ & $N_{\rm meas}$\\
\hline
0.83 &(0,2,0)  &0.69  &0      & 519  & 4152\\
1.25 &(0,2,0)  &0.69  &0      & 1315  & 42080\\
1.67 &(0,2,0)  &0.69  &0      & 1753  & 112192\\
1.25 &(0,2,2)  &1.02  & 1/3   & 417  & 40032\\
1.25 &(0,2,-2) &1.02  & -1/3  & 417  & 40032 \\
\hline
\end{tabular}
\begin{minipage}{9cm}
\caption{Statistics at each $P_3$, $\vec{\Delta}$ and $\xi$. $N_{\rm confs}$ and $N_{\rm meas}$ is the number of analyzed configurations and the total measurements. }
\label{tab:stat_GPDs}
\end{minipage}
\end{center}
\end{minipage}}
\end{table}
\begin{figure*}[htp!]
  \centering
  \subfigure[$F_{H}$ (squares), $F_{\tilde{H}}$ (diamonds) and $F_{H_T}$ (circles).]{\includegraphics[width=0.49\linewidth]{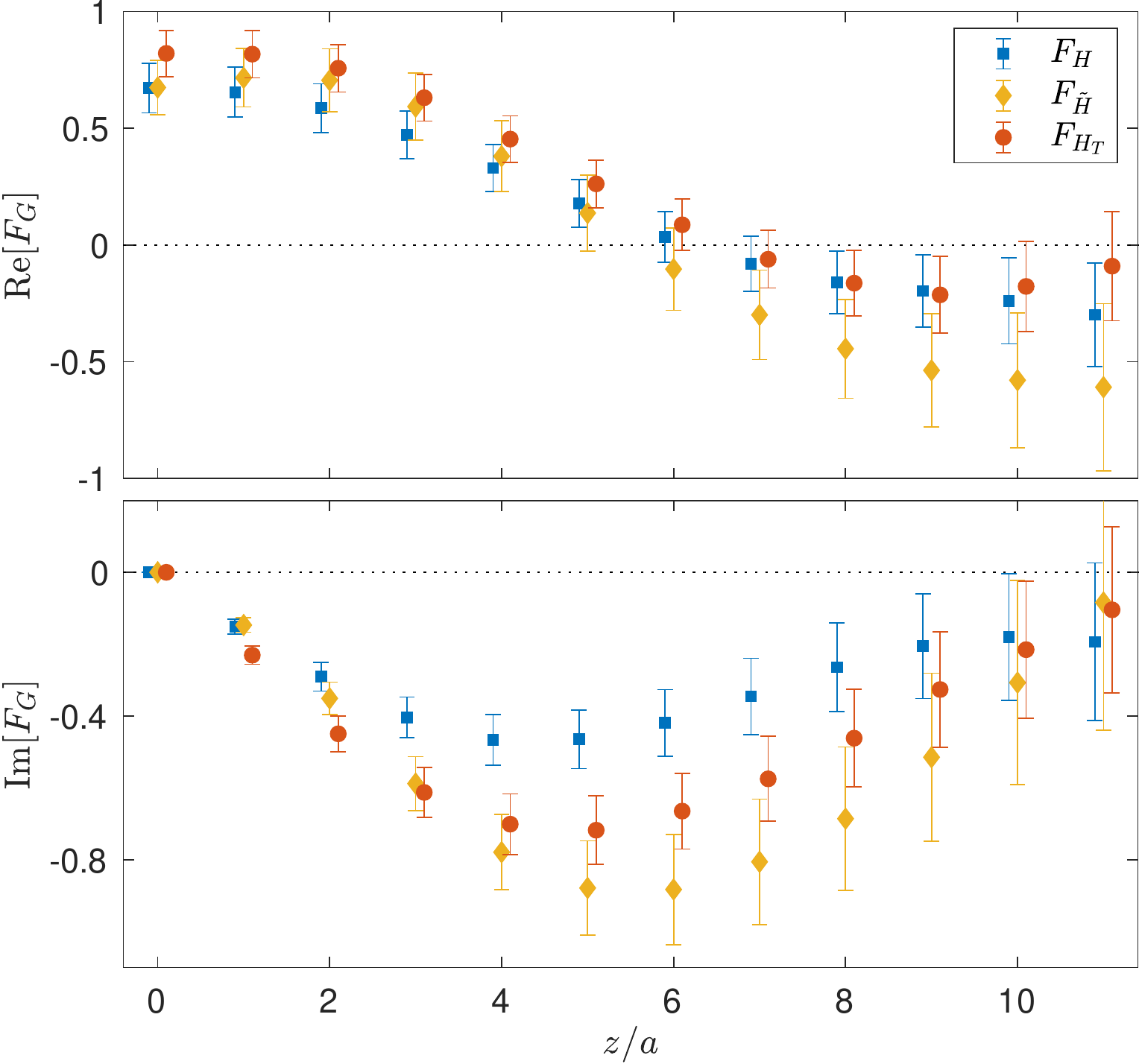}}\hspace{0.4cm}
\subfigure[$F_{E}$ (squares), $F_{E_T}$ (diamonds), $F_{\tilde{E}_T}$ (circles), $F_{\tilde{H}_T}$ (triangles).]{\includegraphics[width=0.48\linewidth]{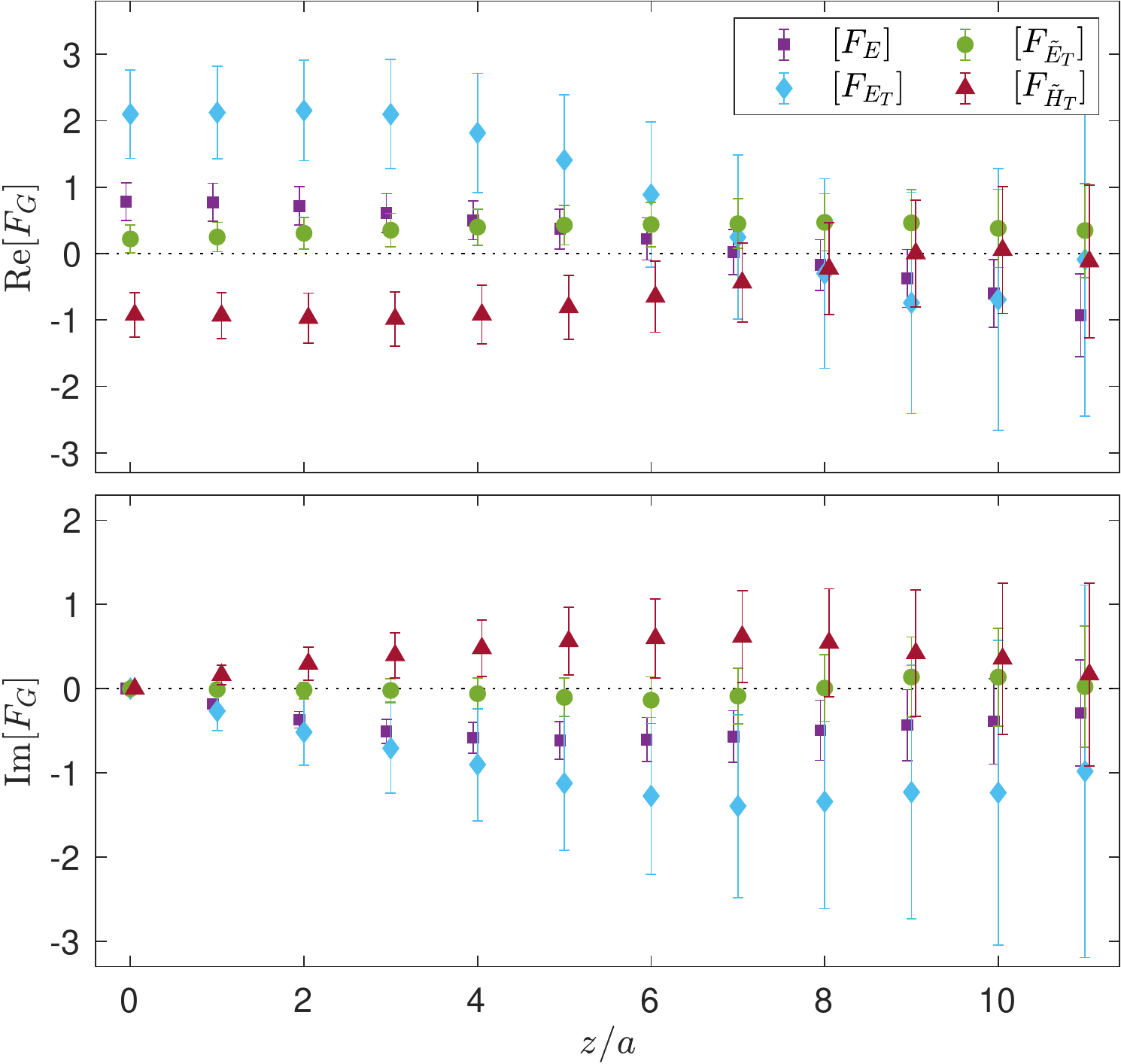}}
    \caption{Results for the matrix elements of the GPDs as functions of $z/a$. The data are in the RI-MOM scheme at $(a\mu_0)^2\approx 2.57$, $P_3=1.67$~GeV and $\xi=0$. Top: real part; bottom: imaginary part.}
    \label{fig:FG}
\end{figure*}

With this setup we compute two independent matrix elements having a vector and axial-vector insertion operator and we apply a decomposition of the type~\ref{eq:decomposition} to find the functions $F_{H},F_{E}$ and $F_{\tilde{H}},F_{\tilde{E}}$, for the unpolarized and helicity case respectively. For the transversity GPDs we need four additional matrix elements that have a tensor structure in the insertion, since there are four functions, $F_{H_T},F_{E_T},F_{\tilde{H}_T},F_{\tilde{E}_T}$ to be disentangled. 

In Fig.~\ref{fig:FG}, panel $(a)$, we show $F_{H},F_{\tilde{H}}$ and $F_{H_T}$, that are the matrix elements of the GPDs with a nonzero value in the forward limit, at $\xi=0$. The momentum is $P_3=1.67$~GeV. At this value of $P_3$, we find that the real and imaginary parts decay to zero as $z/a$ increases; this is not always guaranteed for the lowest momentum considered here  (see~\cite{Alexandrou:2020zbe,Alexandrou:2021bbo,Alexandrou:2021lyf}). We also observe that $F_H$ has the best signal quality and its imaginary part is smaller in magnitude. At $\xi=0$, we also find that the remaining GPDs are in general affected by increased statistical uncertainties, because of their kinematic. The decomposed functions in coordinate space are shown in Fig.~\ref{fig:FG}, panel $(b)$. The real and imaginary parts of $F_{\tilde{E}_T}$ are found to be compatible with zero, in agreement with $\tilde{E}_T$ expected to be odd under replacement $\xi\rightarrow -\xi$~\cite{Meissner:2007rx,Diehl:2003ny}. We also note that at, $\xi=0$, $F_{\tilde{E}}$ is not accessible by decompositions of the form~\ref{eq:decomposition} because its kinematic factor vanishes and therefore here it is not presented.

It is important to notice that at $z=0$ the functions $F_G$ reduce to the elastic FFs, that in lattice QCD are usually computed using local matrix elements in the rest frame of the final nucleon state. In fact, we find that $F_{E}$, $F_{\tilde{H}}$ and $F_{H_T}$ are compatible with the FFs $F_1$, $g_A$ and $A_{T_{10}}$ computed using a twisted mass ensemble with a similar pion mass~\cite{Alexandrou:2013joa,Alexandrou:2013wka}. This serves as an important check of our calculation.
\section{$x$-dependence of the GPDs}
\begin{figure*}[!htp]
\centering 
{\includegraphics[width=0.485\linewidth]{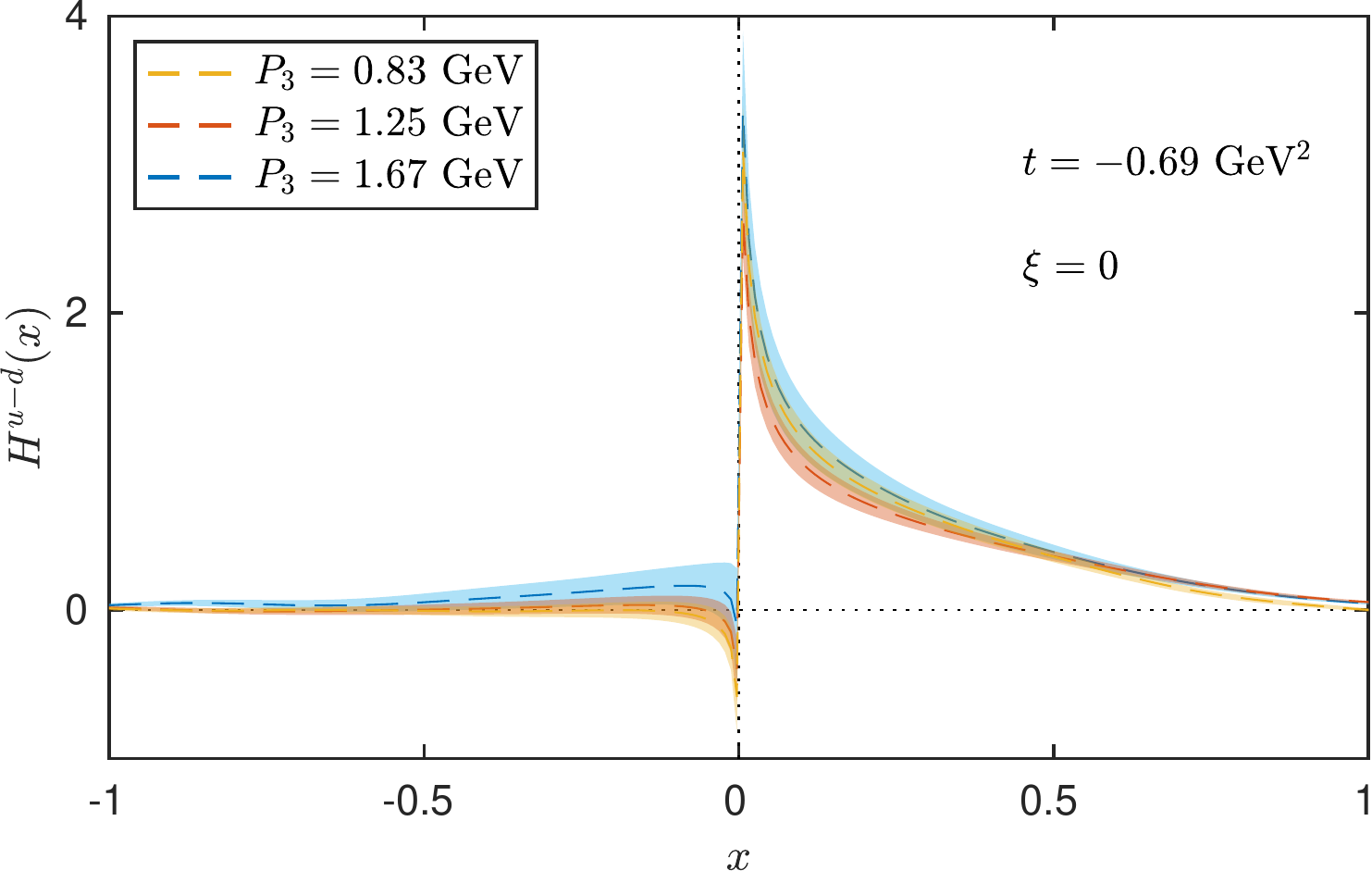}}\hspace{0.1cm}
{\includegraphics[width=0.485\linewidth]{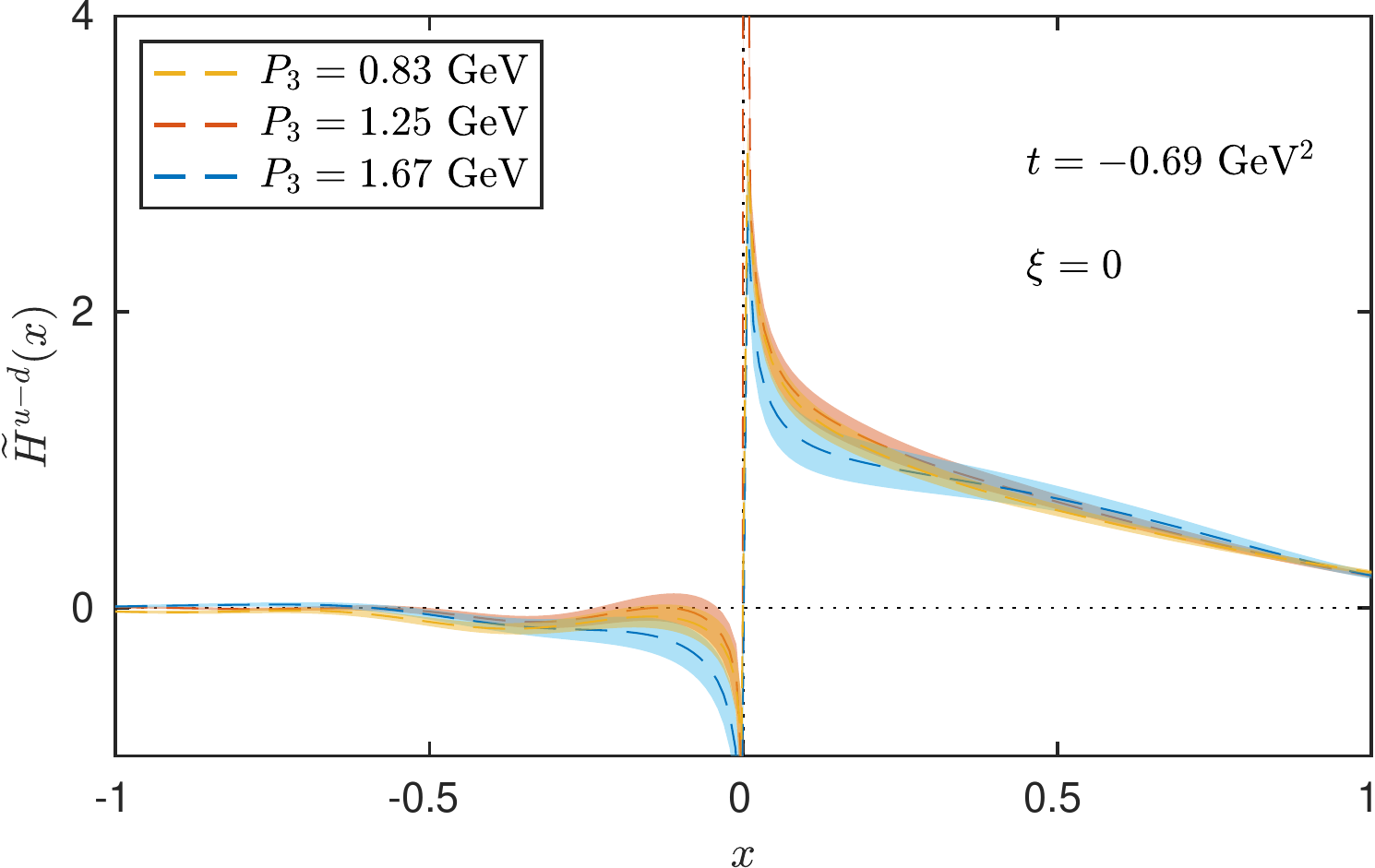}}\hspace{0.1cm}

{\includegraphics[width=0.485\linewidth]{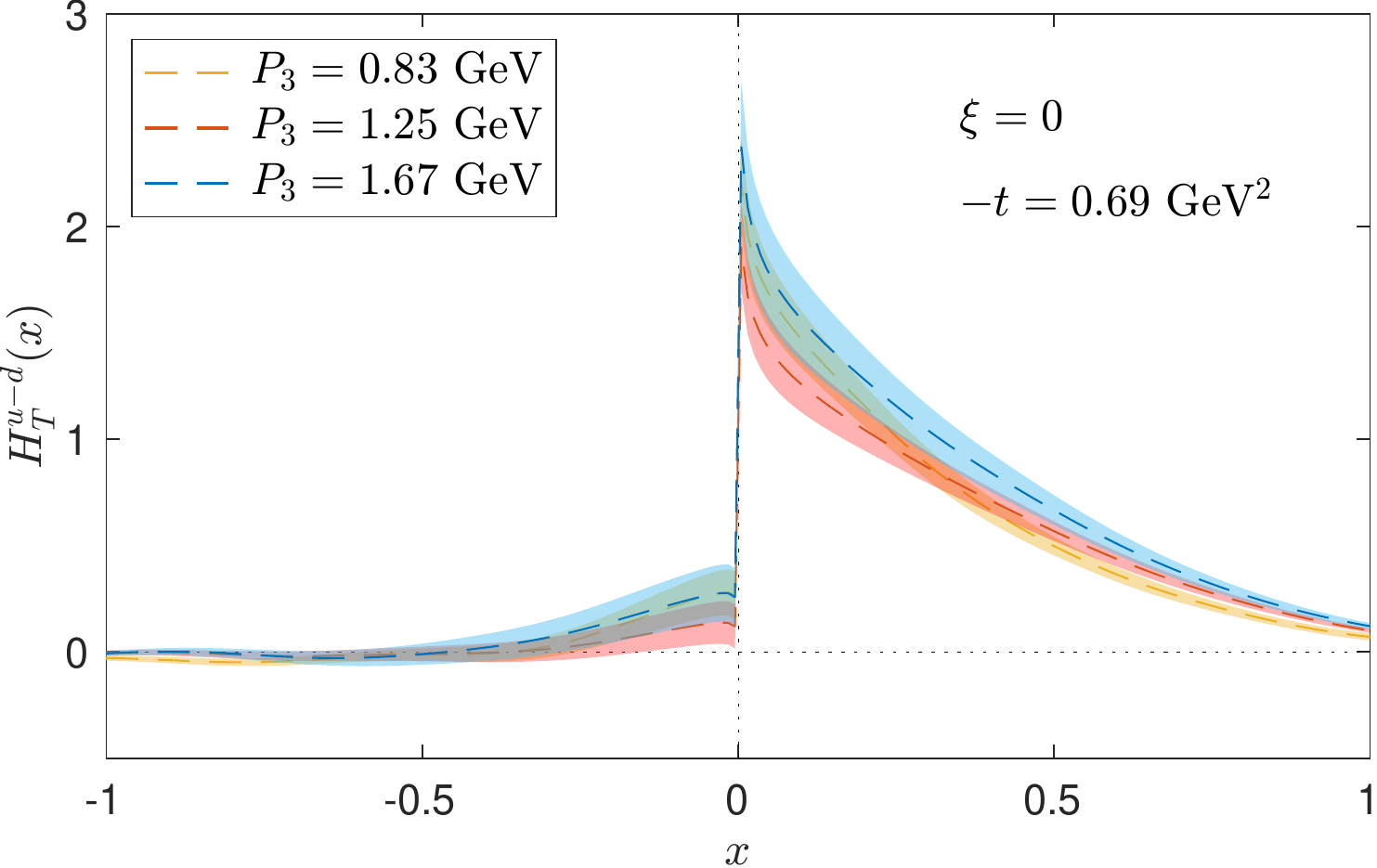}}
\caption{Momentum dependence of the $u-d$ isovector $H$, $\tilde{H}$ and $H_T$ GPDs at $\mu=2$~GeV in the $\MSb$ scheme. Results are shown for $0.83$~GeV (yellow), $1.25$~GeV (red) and $1.67$~GeV (blue). The bands refer only to the statistical uncertainties.}
\label{fig:P_dependence}
\end{figure*}

To obtain the GPDs as a function of $x$, we perform the Fourier transform of the matrix elements $F_G$ to momentum space and apply the matching coefficient of Ref.~\cite{Liu:2019urm}. The final distributions are in the $\MSb$ scheme at a scale $\mu=2$~GeV. The $P_3$-dependence at $\xi=0$ for $H$, $\tilde{H}$ and $H_T$ is shown in Fig.~\ref{fig:P_dependence}, in which the error bands include only statistical uncertainties. At this accuracy we find that the distributions are in agreement for the three nucleon boosts and this holds both for the antiquark $(x<0)$ and quark $(x>0)$ regions. Thus, we decide to focus on the intermediate momentum, $P_3=1.25$~GeV, to study the effect of a nonzero skewness on the lattice GPDs.  

\begin{figure*}[!htp]
{\includegraphics[width=0.485\linewidth]{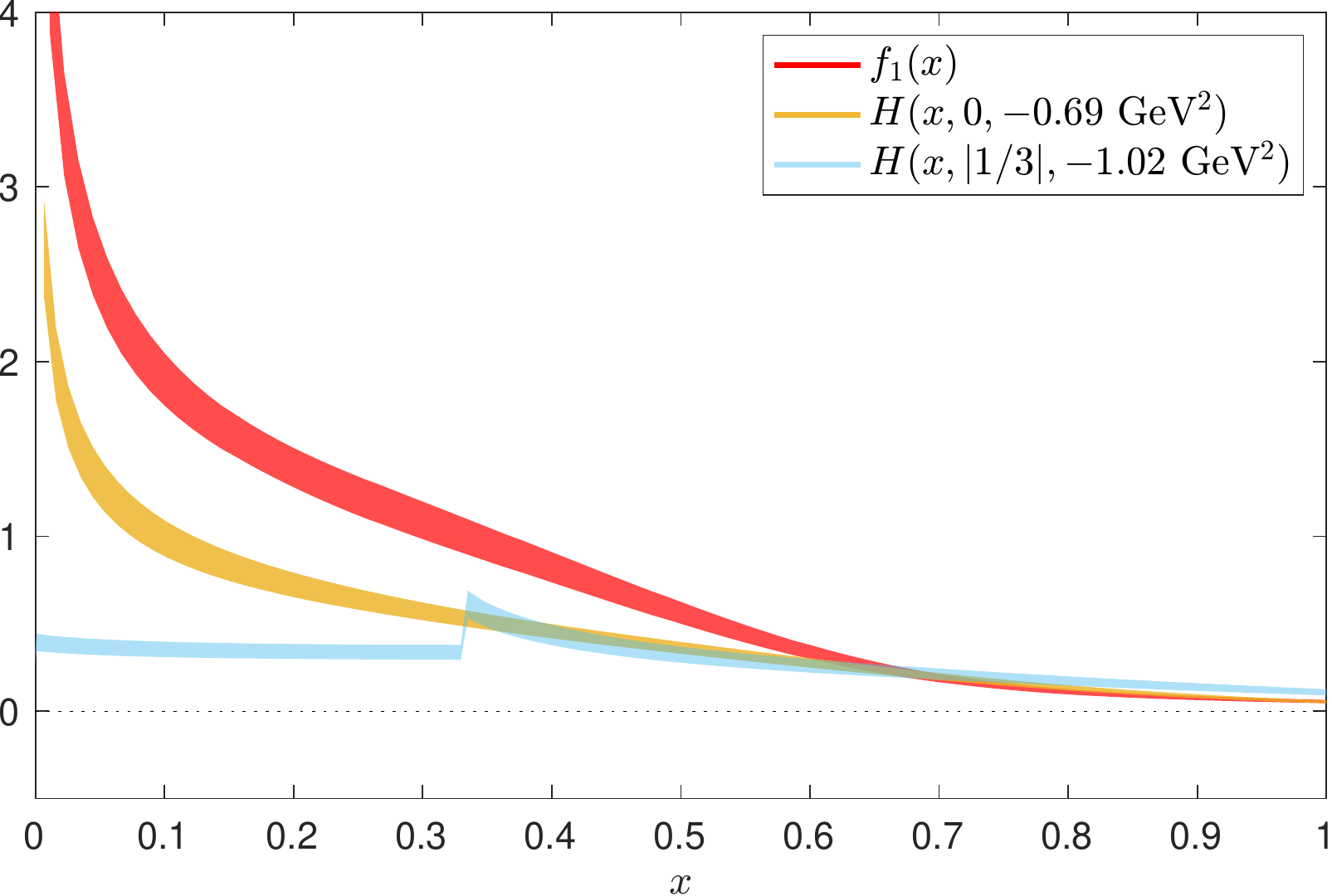}}\hspace{0.3cm}
{\includegraphics[width=0.485\linewidth]{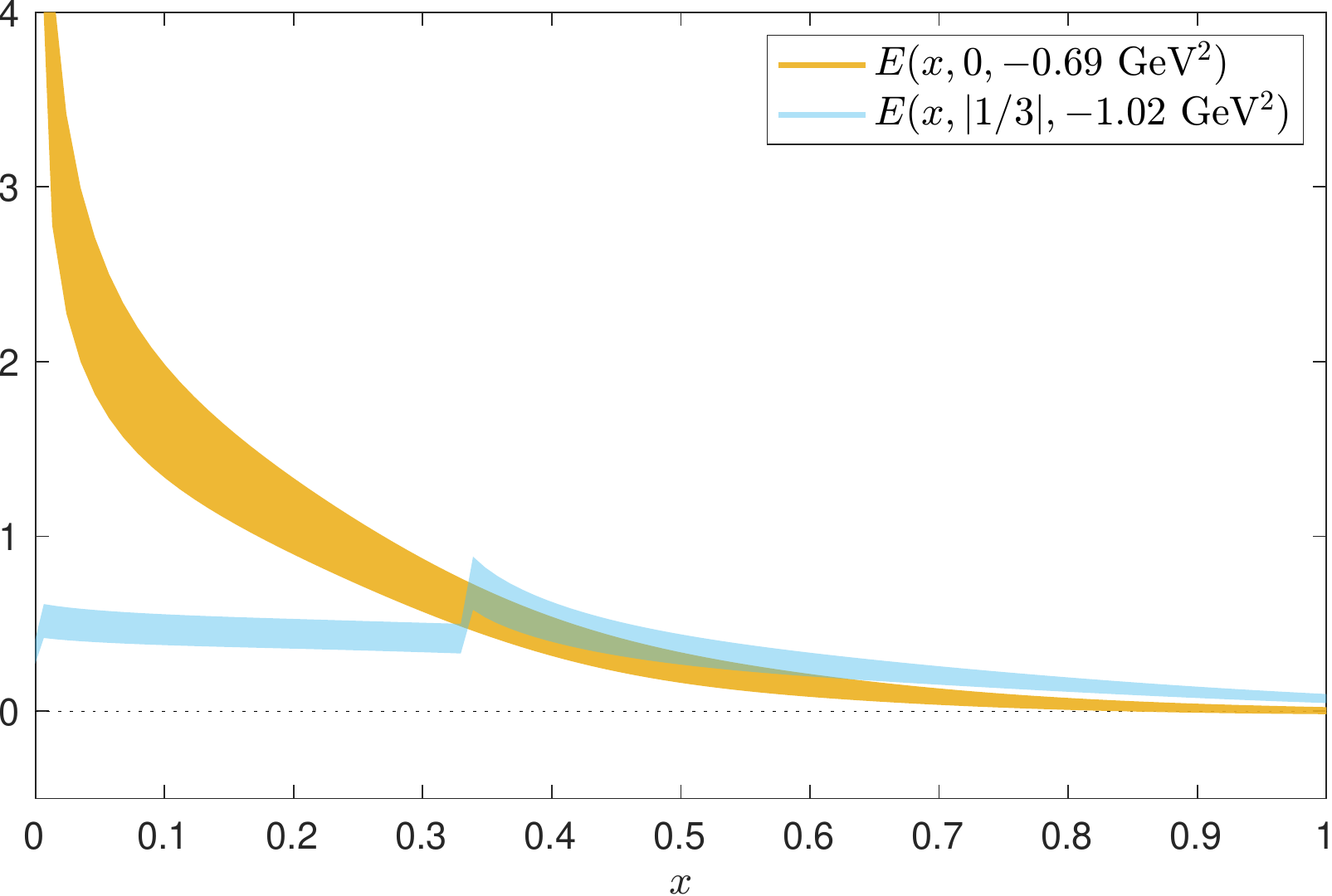}}

{\includegraphics[width=0.485\linewidth]{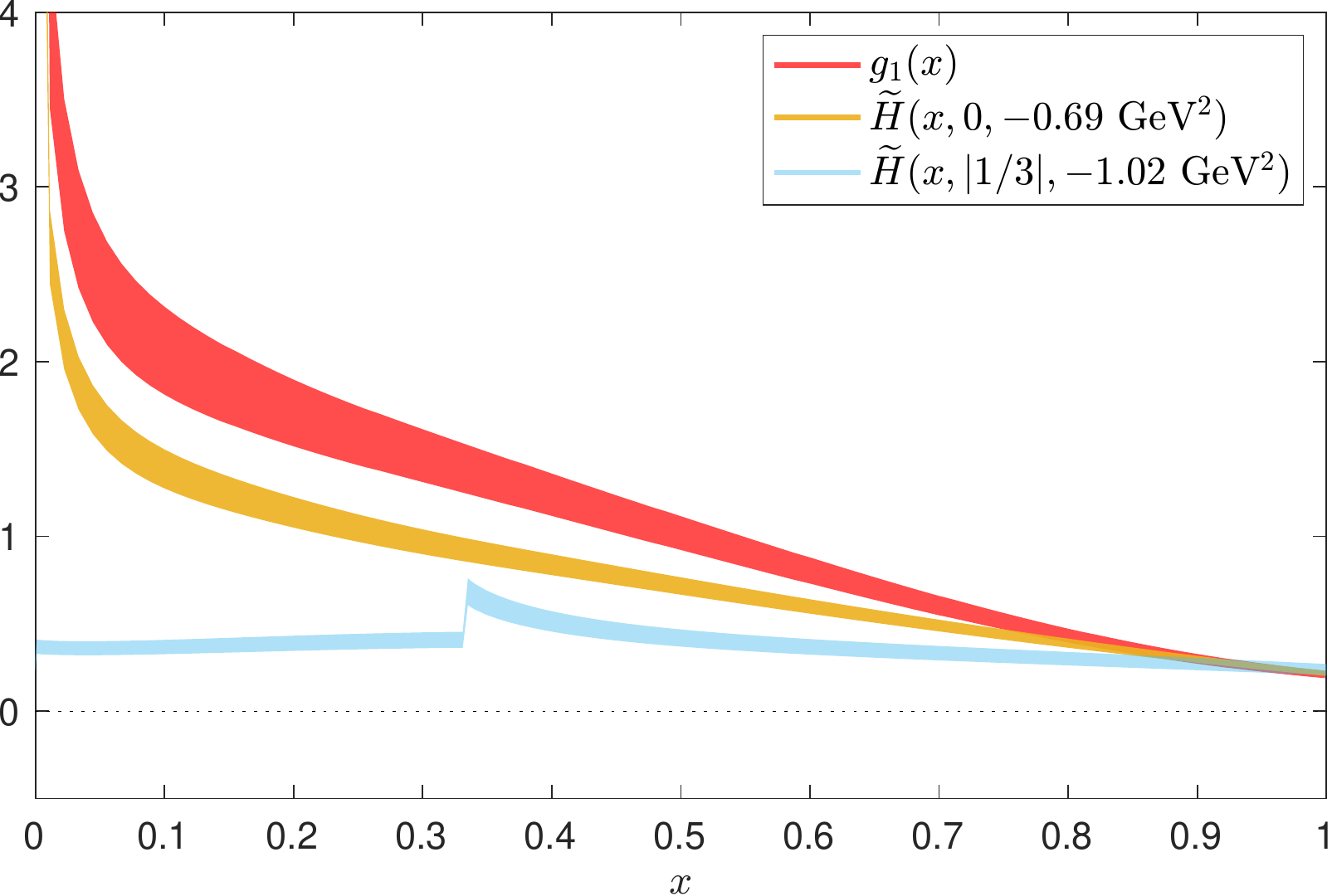}}\hspace{0.3cm}
{\includegraphics[width=0.485\linewidth]{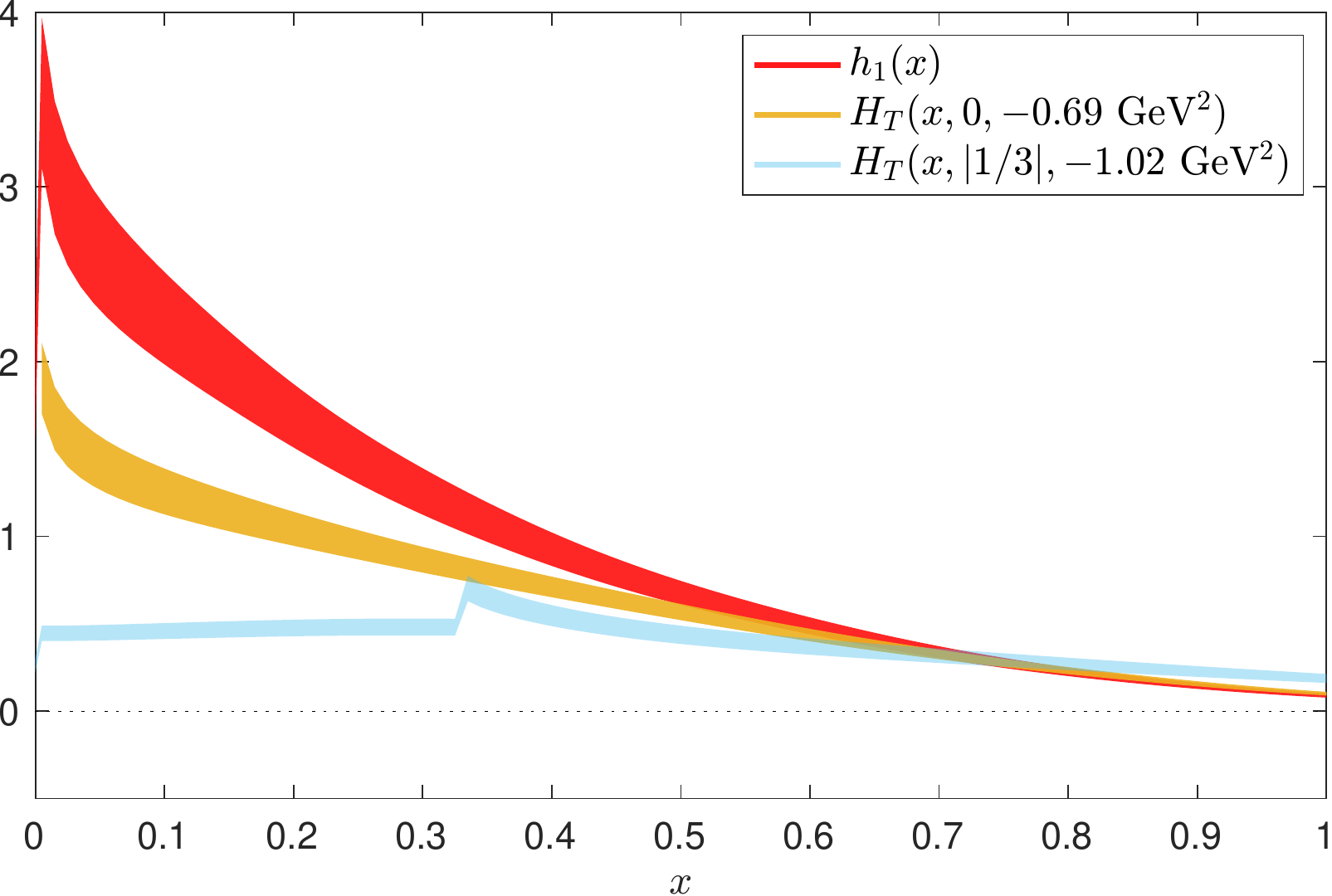}}
\vspace{-0.2cm}
\caption{Comparison between PDFs and GPDs at $P_3=1.25$~GeV, $\xi=0$ and $|\xi|=1/3$ for the unpolarized (top), helicity (bottom left) and transversity (bottom right) case. The PDFs are shown in red, while GPDs at $\lbrace \xi=0,-t=0.69$~GeV$^2\rbrace$ and $\lbrace |\xi|=1/3,-t=1.39$~GeV$^2\rbrace$ are in orange and cyan respectively. The discontinuities at $x=\xi$ are due to uncontrolled higher-twist contaminations that cannot be treated at the present stage.}
\label{fig:nonzeroxi}
\end{figure*}

The GPDs at $\xi\neq 0$ are obtained using a non-vanishing component of the momentum transfer in the boost direction. The value of $\xi$ enters explicitly the matching equations~\cite{Liu:2019urm} that have a different form in the DGLAP~\cite{Dokshitzer:1977sg,Gribov:1972ri,Lipatov:1974qm,Altarelli:1977zs} $(x>|\xi|)$ and in the ERBL~\cite{Efremov:1979qk,Lepage:1980fj} $(x<|\xi|)$ regions. For the ensemble considered in this work the smallest $t$-value with $\xi\neq 0$ is $-t=1.02$~GeV$^2$; this can be obtained either for $\xi=+1/3$ or $\xi=-1/3$. To reduce lattice artifacts we average over these two possibilities. Results for $H$, $E$, $\tilde{H}$ and $H_T$ are shown in Fig.~\ref{fig:nonzeroxi}. The focus is only on the quark region $(x>0)$ because it is less susceptible to systematic errors~\cite{Alexandrou:2020qtt}. We note, once again, that systematic uncertainties are not included yet in the error budget. 
From Fig.~\ref{fig:nonzeroxi} we observe a clear distinction between DGLAP and ERBL regions (cyan bands) at $\xi\neq 0$. In fact, at $x=\xi$, these regions are separated by discontinuities that are however non-physical, as twist-2 GPDs are expected to be continuous functions at the boundaries~\cite{Bhattacharya:2018zxi,Bhattacharya:2019cme}. This effect arises from higher-twist contributions not yet computed in the matching procedure, that contains the leading twist-2 terms. Beside that, we find that GPDs decrease in magnitude as $-t$ increases, as expected from the usual suppression of the elastic FFs with the increase of the momentum transfer. Moreover, such suppression is stronger in the ERBL than in the DGLAP region. When comparing GPDs with PDFs (computed on the same gauge ensemble) we find that the PDFs are dominant for most values of $x$, as expected from the relations $f_1(x)=H(x,0,0)$, $g_1(x)=\tilde{H}(x,0,0)$ and $h_1(x)=H_T(x,0,0)$. At large $x$ the $t$-dependence seems to vanish especially for the unpolarized case, and in the future it would be interesting to study this effect in relation to the work of Ref.~\cite{Yuan:2003fs}, where for $H$ it was found an asymptotic behavior $(1-x)^3/(1-\xi^2)^2$ through power counting analysis.

As a final cross-check, we verify that the integral in $x$ of the GPDs in $[-1,+1]$ is in agreement with the matrix elements $F_G$ at $z=0$ and with the FFs of the work~\cite{Alexandrou:2013joa,Alexandrou:2013wka}. This result is highly non-trivial, as the reconstruction of the $x$-dependence involves many delicate steps with different systematics, such as the Fourier-transform to momentum space and the matching procedure to light-cone GPDs. 

\section{Conclusions and future work}
In this manuscript we report on our first effort to computing the isovector GPDs of the proton from lattice QCD. We use a single ensemble of twisted mass fermions with pion mass $M_\pi=260$~MeV and study the momentum dependence of the GPDs using a nucleon boost up to 1.67~GeV. Convergence is seen between $P_3=1.25$ and $1.67$ GeV, for the unpolarized and the polarized GPDs (see also Refs.~\cite{Alexandrou:2020zbe,Alexandrou:2021bbo,Alexandrou:2021lyf}). The GPDs are extracted at $-t=0.69$~GeV$^2,\xi=0$ and $-t=1.02$~GeV$^2,|\xi|=1/3$ and they show the expected behavior, i.e. they are suppressed as the momentum transfer increases. In addition, $H$, $\tilde{H}$ and $H_T$ have statistical uncertainties that are similar to the corresponding PDFs. In the future, we plan to include additional values of the momentum transfer, for which larger volume ensembles will be crucial. The final goal is to extract, among others, the impact parameter distributions via a Fourier-transform in the $t$-space and the quark orbital angular momentum directly from the GPDs. Eventually, as for any lattice calculation, dedicated studies are necessary to estimate various systematic uncertainties due to the pion mass, lattice spacing and finite volume effects.
\section{Acknowledgements}
A.S. thanks the organizers of the 19th HADRON Conference for the invitation to present this work. K.C.\ is supported by the National Science Centre (Poland) grant SONATA BIS no.\ 2016/22/E/ST2/00013. M.C. and A.S. acknowledge financial support by the U.S. Department of Energy Early Career Award under Grant No.\ DE-SC0020405. K.H. is supported by the Cyprus Research and Innovation Foundation under grant POST-DOC/0718/0100. 
F.S.\ was funded by the NSFC and the Deutsche Forschungsgemeinschaft (DFG, German Research
Foundation) through the funds provided to the Sino-German Collaborative Research Center TRR110 ``Symmetries and the Emergence of Structure in QCD'' (NSFC Grant No. 12070131001, DFG Project-ID 196253076 - TRR 110).
Partial support is provided by the European Joint Doctorate program STIMULATE of the European Union's Horizon 2020 research and innovation programme under grant agreement No. 765048. Computations for this work were carried out in part on facilities of the USQCD Collaboration, which are funded by the Office of Science of the U.S. Department of Energy. This research was supported in part by PLGrid Infrastructure (Prometheus supercomputer at AGH Cyfronet in Cracow). 
Computations were also partially performed at the Poznan Supercomputing and Networking Center (Eagle supercomputer), the Interdisciplinary Centre for Mathematical and Computational Modelling of the Warsaw University (Okeanos supercomputer) and at the Academic Computer Centre in Gda\'nsk (Tryton supercomputer). The gauge configurations have been generated by the Extended Twisted Mass Collaboration on the KNL (A2) Partition of Marconi at CINECA, through the Prace project Pra13\_3304 ``SIMPHYS".

\end{multicols}
\medline
\begin{multicols}{2}

\end{multicols}
\end{document}